\documentclass[a4paper,twocolumn,prb]{revtex4}
\usepackage{amsmath}
\usepackage{graphicx}
\usepackage{color}
\usepackage{subfigure}
\usepackage{hyperref}
\begin{document}
\title{Chaos in Sandpile Models}
\author{Saman Moghimi-Araghi}
\affiliation{Physics department, Sharif University of Technology, P.O. Box 11155-9161, Tehran, Iran}
\author{Ali Mollabashi}
\affiliation{Physics department, Sharif University of Technology, P.O. Box 11155-9161, Tehran, Iran}
\begin{abstract}
We have investigated the "weak chaos" exponent to see if it can be considered as a classification parameter of different sandpile models. Simulation results show that "weak chaos" exponent may be one of the characteristic exponents of the attractor of \textit{deterministic} models. We have shown that the (Abelian) BTW sandpile model and the (non Abelian) Zhang model posses different "weak chaos" exponents, so they may belong to different universality classes. We have also shown that \textit{stochasticity} destroys "weak chaos" exponents' effectiveness so it slows down the divergence of nearby configurations. Finally we show that getting off the critical point destroys this behavior of deterministic models.
\end{abstract}
\maketitle
\section{Introduction}
Bak, Tang, and Wissenfeld (BTW) introduced the concept of self-organized criticality (SOC) and the so-called BTW sandpile model as a description of power spacial and temporal correlations observed in a wide range of natural phenomenon \cite{BTW87,BTW88}. During the past two decades, more sandpile models were introduced by different variations of the main paradigm of SOC, the BTW model, in order to gain more realistic models. These models differ in some properties such as: discrete or continuous height variable, Abelian or non Abelian toppling rule, stochastic or deterministic toppling rule, directed or non directed (or even non directed on average) toppling current, sticky or non-sticky grains, and etc. We now have a large number of different sandpile-like models each model having it's own set of critical exponents. The number of these models are exceeding but after about two decades, a little is known about their universal classification. Although some numerical studies are done to address the universality of the critical behavior of different sandpile models but these studies are in contradiction with each other. For example Manna's classification \cite{M91} is in contrast with Ben-Hur and Biham's \cite{Universality96} one, and the latter is in contradiction with Chessa \textit{et al.} \cite{CSVZ99} classification. Also more recently fixed energy sandpiles (FES) are introduced in order to study the critical behavior of SOC and classifying sandpile models which has not gained any serious success yet and even it does not seem to be a successful career (for example see \cite{P10}).

The first classification was done by Manna. He put his model and the BTW model in the same universality class ascribing the observed difference in critical exponents to finite size effects \cite{M91}. This result was verified by Grassberger and Manna \cite{GSSM91,SSM91}. Another effort was done by Díaz-Guilera and Corral based on renormalization group which resulted in classifying BTW and Zhang model in the same class \cite{DG94,CDG97} (which confirmed Zhang's conjecture \cite{Z89}). Ben-Hur and Biham \cite{Universality96} studied the most complete set of critical exponents $\{s, a, t, d, r, p\}$, based on the evolution of conditional expectation values (See Christensen \textit{et al.} \cite{CFJ91}). These exponents are related to: (s) the size of an avalanche, (a) the area of an avalanche, (t) the time duration of an avalanche, (d) the maximal distance between the origin and the sites that an avalanche cluster touches, (r) the radius of gyration of an avalanche cluster, and (p) the perimeter of an avalanche cluster. They showed that sandpile models are classified in three groups of \textit{non directed}, \textit{non directed on average}, and \textit{directed} models \cite{Universality96}. As a result of their study, BTW and Zhang models belong to the same universality class (non directed) which Manna model (as a non directed on average model) does not and directed models belong to another class. Chessa \textit{et al.} made some systematic corrections on Ben-Hur and Biham's method and claimed that both stochastic and deterministic sandpile models belong to the same universality class \cite{CSVZ99}.    

On the other hand, Bak and Chen had investigated the chaotic behavior of a block-spring model (which was introduced for simulating earthquake dynamics) \cite{BC90}. They have shown that although the largest Lyapanuv exponent of this model is zero but nearby configurations separate in a power-law manner and they called it "weak chaos" \cite{BC90}. Based on this study, Bak, Tang, and Chen (BTC) conjectured that SOC takes the system to the border of chaos. They also argued that this behavior is not because of exponential sensitivity to initial conditions but the critical fluctuations of the system. They conjectured that in this manner, "weak chaos" is another aspect of the criticality of the attractor so the "weak chaos" exponent \textit{is} a characteristic exponent of the system \cite{BC90, BT89}.

Although Vieira and Lichtenberg found a counter example for BTC's conjecture \cite{VL96}, we have numerically checked it for different sandpile models. We have shown that BTC's conjecture is truly verified in BTW (also in CBTW) and Zhang models but it's not true in Manna stochastic model. Dhar-Ramaswamy directed model behaves more complicated with at least two different regimes. 

In this paper we first define some sandpile models and discuss the time evolution of nearby points in their configuration space. We will finally discuss the same behavior in the off critical regime. 
\section{Defining the Models}
We have studied sandpile models on two dimensional square, triangular, and honeycomb lattices. A height variable is assigned to each site $(i, j)$ of the lattice which can be discrete or continuous depending on the model. This height variable could be interpreted as "energy". At each time step, sand is added to a randomly chosen site. Whenever the height of a site $z_{i,j}$, exceeds the critical height $(z_{i,j}>z_c)$, the site would relax through the related toppling rule. Relaxation of a site would cause other sites to become unstable so they would topple and a chain reaction called an avalanche continues until all sites become stable. The rate of energy injection is so slow that an unstable configuration will relax before next grain is added.
\begin{figure}
\centerline{\includegraphics[scale=.65]{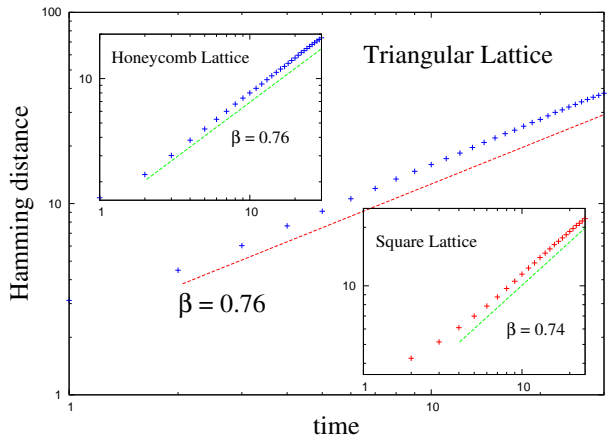}}
\caption{Evolution of $H(t)$ for BTW model on three different lattices on Log scale for $L=256$ and 10000 samples. The down-right, up-left, and the main graphs correspond to square, honeycomb, and triangular lattices correspondingly. Lattice effects which slows down $H(t)$ in the first few time steps is mostly seen on square lattice.}
\label{fig:BTW}
\end{figure}
\\\textbf{BTW model:} the height variable of this model is discrete and its critical height is 4, 6, and 3 on square, triangular, and honeycomb lattices correspondingly. When $z_{i,j}>z_c$ the $(i, j)$th site relaxes: 
\begin{equation}
\begin{array}{l}
z_{i,j} \rightarrow z_{i,j}-z_c \\
z_{nn} \rightarrow z_{nn}+1
\end{array}
\end{equation}
which $nn$ means the nearest neighbors. The boundary condition of this model is chosen to be open on all boundaries.
\\\textbf{Zhang model:} the height variable in this model is continuous and its critical height can always be taken to be 1. At each time step a random amount of sand $\delta$, which we take it to be in the set $(0,0.25)$, is added to a randomly chosen site. If $z_{i,j}>1$ this site topples by:
\begin{equation}
\begin{array}{l}
z_{i,j} \rightarrow 0 \\
z_{nn} \rightarrow z_{nn}+\frac{z_{i,j}}{nnn}
\end{array}
\end{equation}
$nnn$ is the number of nearest neighbors. Boundary conditions in this model is also open on all boundaries. 
\\\textbf{Continuous BTW model (CBTW):} this model is a continuous version of BTW model \cite{ALM07}. Again as its height variable is continuous, the critical height can be taken to be 1. At each time step an amount of sand $\delta$, with $0<\delta<0.25$, is added to a randomly chosen site, and the toppling rule is given via
\begin{equation}
\begin{array}{l}
z_{i,j} \rightarrow z_{i,j}-1 \\
z_{nn} \rightarrow z_{nn}+\frac{1}{nnn}
\end{array}
\end{equation}
$nnn$ is again the number of nearest neighbors.
\\\textbf{Manna model:} in this stochastic model the critical height is taken to be 2 on square and triangular lattices. On the square lattice when an unstable site is going to topple, it either gives its left and right or its up and down neighbors one unit of sand each with equal probability. On the triangular lattice when a site topples, it gives one grain of sand to each facing neighbors with a probability of $1/3$. The boundary conditions on this model is chosen to be open on all boundaries.
\\\textbf{Directed model:} Dhar-Ramaswamy model \cite{DR89} is referred as the directed model. This model is defined on a square lattice in the $(1,1)$ direction. The critical height is 2 and when a site topples it gives one sand to each two \textit{down} neighbors. The vertical boundary condition is cylindrical, sands are added to the system from the most top row and leave it from the lowest row.
\begin{figure}
\centerline{\includegraphics[scale=.65]{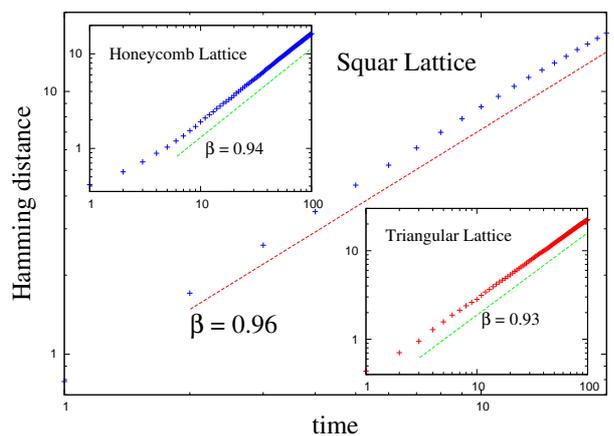}}
\caption{Evolution of $H(t)$ for Zhang model on three different lattices on Log scale for $L=256$ and 10000 samples. The main, down-right, and the up-left graphs correspond to square, triangular, and honeycomb lattices correspondingly. Lattice effects which slows down $H(t)$ in the first few time steps is mostly seen on honeycomb lattice.}
\label{fig:Zhang}
\end{figure}
\section{Chaos in Sandpile Models}
To study chaos in different sandpile models we have monitored the time evolution of the distance of two nearby configurations. The Hamming distance of two configurations at time $t$ which is defined by
\begin{equation}
H(t)=\Big\{ \sum_{i,j=1}^{L}[z_{i,j}(t)-z'_{i,j}(t)]^{2}\Big\} ^{\frac{1}{2}}
\end{equation}
is used as the distance of two sandpiles \cite{ES91}. We have first provided a configuration $\mathcal{Z}$ which its height variables are denoted by $\{z_{i,j}\}$, then we manipulate the heights of about $0.0005$ of the whole sites chosen randomly, to prepare a nearby configuration 
$\mathcal{Z'}$ with heights $\{z'_{i,j}\}$. In each time step an amount of sand is added to a \textit{similar} randomly chosen site of $\mathcal{Z}$ and $\mathcal{Z'}$ and $H(t)$ is calculated after relaxation of both sandpiles.
\begin{figure}
\centerline{\includegraphics[scale=.65]{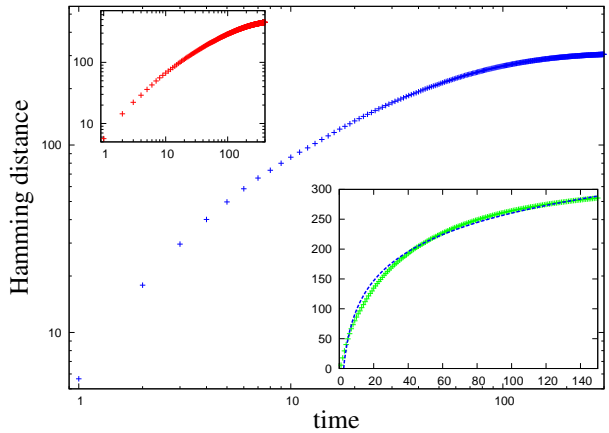}}
\caption{Evolution of $H(t)$ for Manna model on square and triangular lattices on Log scale for $L=256$ and 5000 samples. The main and up-left graphs correspond to square and triangular lattices correspondingly. The down-right graph shows a logarithmic fit for   this model on square lattice.}
\label{fig:Manna}
\end{figure}
\\\textbf{BTW model:} figure \ref{fig:BTW} shows the evolution of $H(t)$ for this model on square, triangular, and honeycomb lattices. Comparing $H(t)$ on different lattices shows that (ignoring a few first time steps which is attributed to lattice effects)
\begin{equation}\label{eq:weakChaos}
H(t)\sim t^{\beta},
\end{equation}
with $\beta_{square}=0.74\pm 0.03$, $\beta_{triangular}=0.76\pm 0.03$, and $\beta_{honeycomb}=0.76\pm 0.03$.

Simulations of this model is done for $L=64, 100, 128, 200, 256, 300, 400, 512, 1024$ and for $L \gtrsim 200$ the exponents are independent of $L$ (for smaller sandpiles the exponents increase slightly by system size). As it's seen in fig \ref{fig:BTW} and table \ref{tbl}, the "weak chaos" exponent does not depend on lattice geometry. This is a good evidence for for considering it as universal property of this model.

The behavior of CBTW model is exactly the same as this model as expected.
\\\textbf{Zhang model:} figure \ref{fig:Zhang} shows the evolution of $H(t)$ for this model again on square, triangular, and honeycomb lattices. Comparing these three shows that (ignoring the first few points) Zhang model obeys BTC's conjecture with exponents $\beta_{square}=0.96\pm 0.05$, $\beta_{triangular}=0.93\pm 0.05$, and $\beta_{honeycomb}=0.94\pm 0.05$. Again the exponents are not lattice dependent.
\begin{figure}
\centerline{\includegraphics[scale=.65]{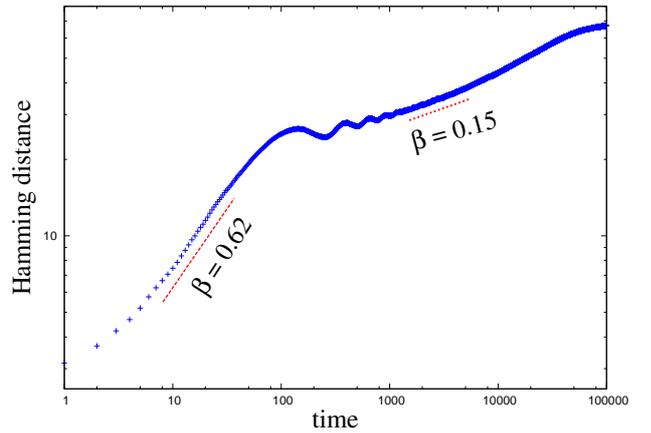}}
\caption{Evolution of Hamming distance for Dhar-Ramaswamy directed model for $L=128$ and 2000 samples. This model shows a more complicated behavior. There are two power-law regimes with $0.64\pm0.04$ and $0.15\pm0.03$ correspondingly}
\label{fig:DASM}
\end{figure}
\begin{table}
\begin{center}
\begin{tabular}{c c c c}
\toprule
Model & Lattice type & size & $\beta$
\\
\hline
BTW & square & 256 & $0.74\pm 0.03$\\
BTW & triangular & 256 & $0.76\pm 0.03$\\
BTW & honeycomb & 256 & $0.76\pm 0.03$\\
\hline
Zhang & square & 256 & $0.96\pm 0.05$\\
Zhang & triangular & 256 & $0.93\pm 0.05$\\
Zhang & honeycomb & 256 & $0.94\pm 0.05$\\
\hline
\end{tabular}
\caption{Weak chaos exponent for BTW and Zhang models on square, triangular, and honeycomb lattices.Since lattice effects can not be separated from power-law regime in a standard way, the errors are reported on the base of simulating each model five times each time containing 5000 samples.}
\label{tbl}
\end{center}
\end{table}
\\\textbf{Manna model:} as it's seen in figure \ref{fig:Manna} the evolution of $H(t)$ in this model does not obey eq. \ref{eq:weakChaos} on square lattice rather it's something like 
\begin{equation}\label{eq:log}
H(t)\sim \log t.
\end{equation}
This \textit{very weak chaos} behavior, if one calls that so, is also confirmed on triangular (fig \ref{fig:Manna}) and rhombic lattices which shows that Manna's attractor is not alike BTW one, and they may not be classified in the same universality class.
\\\textbf{Dhar-Ramaswamy model:} figure shows the evolution of $H(t)$ for this directed model. The few first points are attributed to lattice effects (as it's seen in other models on different lattices too) and there is two different regimes of power behavior after which $H(t)$ saturates because of finite size effects; the exponents are $0.64\pm0.04$ and $0.15\pm0.03$ correspondingly. The intermediate regime is unknown to us.

We have also studied a mixed Manna-Zhang model. In this model when a site exceeds the critical height, either the left and right or up and down neighbors are each given half of its energy by equal probability. Deviation from "weak chaos" behavior is also seen in this model. It seems that the stochastic property is responsible for transition from "weak chaos" to "very weak chaos" behavior.

Since "weak chaos" is not observed in all sandpile models, particularly in Manna model, self-organized criticality is not necessarily accompanied by "weak chaos". Therefore BTC's conjecture seems not to be correct in general. 
Since the "weak chaos" exponent of BTW and Zhang deterministic models do not depend on the lattice geometry, it may be considered as a universal property of these models and it may be viewed as a characteristic exponent (table \ref{tbl}). If so, since BTW's weak chaos exponent is $0.74\pm0.03$ and Zhang's weak chaos exponent is $0.96\pm0.05$ these two models (as an Abelian model and a non Abelian model) can't be classified in the same universality class. In this manner our simulation based results do not agree with Zhang's conjecture about the unification of BTW model and his model in the thermodynamic limit \cite{Z89}, Ben-Hur and Biham's classification \cite{Universality96}, and Díaz-Guilera and Corral's classification \cite{DG94,CDG97}. It should be noted that these classifications are all based on \textit{static} considerations, where we have considered a \textit{dynamic} property of the attractor which might cause this disagreement.

On the other hand, although the root of this behavior of Manna model is not known, this classification based on the "weak chaos" exponent may put Manna model in a different class from BTW and Zhang models which is consistent from this aspect with Ben-Hur and Biham's classification \cite{Universality96} but inconsistent with Manna's \cite{M91,GSSM91,SSM91} and Chessa \textit{et al.} \cite{CSVZ99} results.
\begin{figure}
\centerline{\includegraphics[scale=.65]{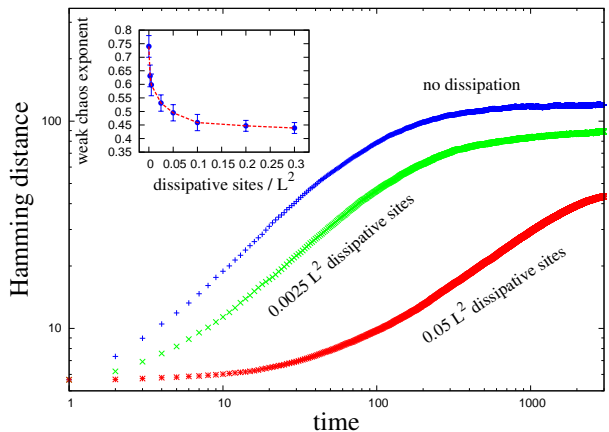}}
\caption{Evolution of $H(t)$ for dissipative BTW model no square lattice with $L=256$ for 5000 samples. At large times $H(t)$ saturates because of finite size effects. The blue graph represents BTW without bulk dissipation. The green and red graphs correspond to $164(=0.0025 L^2)$ and $3277(=0.05 L^2)$ dissipative sites with $k=1$. The up-left graph shows how $\beta$ decreases rapidly by increase of dissipation amount (for fixed $k(=1)$). The asymptotic value of $\beta$ is $0.43$.}
\label{fig:disBTW}
\end{figure}
\section{dissipative models}
All we have reported above is related to sandpile models at the critical point. What happens to this behavior getting off the critical point? Do they still show weak chaos behavior? (It should be noted that since we have not studied fixed energy sandpile models, off critical states here means those states which their mean energy $\rho$ satisfies $\rho<\rho_c$, where $\rho_c$ is the critical mean energy.)

Adding bulk dissipation to sandpile models can be done using different methods. For discrete models the toppling rule is changed in some randomly chosen sites called \textit{dissipative sites}. If the $(i,j)$th site is a dissipative one, it topples when $z_{i,j}>z_c+k$ where $k$ is a positive integer. In this method dissipation is controlled by two parameters, the number of dissipative sites and $k$. In continuous models there is no need of selected dissipative sites and we can impose the so called new toppling rule to all sites where $k$ this time is a real number, therefore dissipation is controlled only by $k$. In discrete models we can also use their continuous version and add arbitrary dissipation to all sites not to impose another stochastic parameter in the system. We have used the first method for the discrete BTW model and the second method for (continuous) Zhang model.

Although bulk dissipation does not differ in principle from boundary dissipation, its importance is because it imposes a characteristic length in the system which destroys criticality (for example see \cite{MKK90}). Since bulk dissipation delays some toppling events we expect the rate of divergence of nearby configurations to decrease when dissipation increases. So we do not expect off critical BTW and Zhang models to show "weak chaos" behavior.
\begin{figure}
\centerline{\includegraphics[scale=.65]{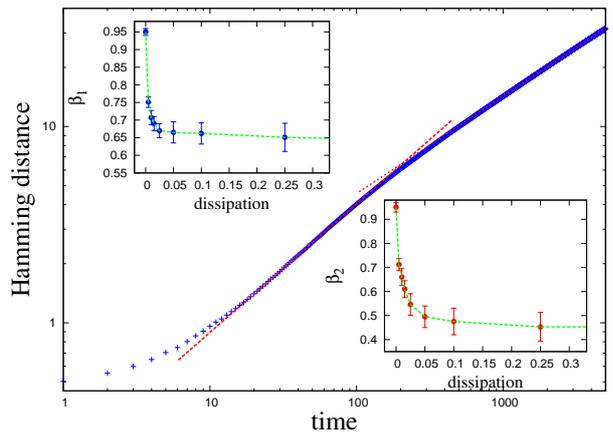}}
\caption{Evolution $H(t)$ for dissipative Zhang model on square lattice for $L=256$ and 10000 samples. The critical height is one and each site dissipates $0.5$ unit of sand. In the main graph $\beta_1=0.64$ and $\beta_2=0.47$. The up-left and down-right graphs show the rapid decrease of both $\beta_1$ and $\beta_2$ by increase of dissipation amount. The asymptotic values are $0.62$ and $0.42$ correspondingly}
\label{fig:disZhang}
\end{figure}

Fig \ref{fig:disBTW} shows that in BTW model by increase of dissipation, despite the interval of lattice effects which takes more long, the exponent of weak chaos region $\beta$ decreases rapidly. This is exactly what we expect because the characteristic length rapidly becomes comparable with the system size, therefore "weak chaos" behavior decays. The up-left graph of figure \ref{fig:disBTW} shows how rapid decrease of the exponent versus dissipation magnitude in BTW model. 

Fig \ref{fig:disZhang} shows that Zhang model does not behave like BTW. The main graph shows the evolution of $H(t)$ for $0.5$ sand unit dissipation at each site. Ascribing first few steps to lattice effects, there are to regimes of power-law behavior in this model which both exponents $\beta_1$ and $\beta_2$ decrease rapidly as the magnitude of dissipation (the characteristic length of the system) increases. These are shown in up-left and down-right graphs of fig \ref{fig:disZhang}. Why two different regimes appear is steel unknown to us.

\section{conclusions}
Our simulation results show that both BTW and Zhang models obey BTC's conjecture except some lattice effects where Manna model does not because of its stochastic property. Dhar-Ramaswamy directed model shows a more complicated behavior where it shows two independent regimes of power-law with an unknown regime in between. So our results contain both examples and counter-examples of BTC's previous conjecture. Although weak chaos exponent does not seem to be a general characteristic of sandpile models but we have found a good evidence to concern it as a test for different universal classifications of these models. By means of this test (Abelian) BTW, (non-Abelian) Zhang, stochastic Manna, and directed Dhar-Ramaswamy models all seem to belong to different universality classes (where non of the offered classifications are in a complete accordance with). We have also shown that as we tend to off critical states, weak chaos behavior seems to disappear were in Zhang model an unknown split of $H(t)$ into two power-law regimes is seen.
\section{Acknowledgment}
We would like to thank S. Rouhani for his helpful comments and careful reading of the manuscript.

\end{document}